\documentclass[aps,prl,twocolumn,groupedaddress]{revtex4}
\usepackage{hyperref} 

\begin{document}


\title{The Transition Amplitude for 2T Physics}


\author{Jo\~ao E. Frederico} \email{joaoeduardo@ufmt.br}
\affiliation{Universidade Federal de Mato Grosso, Deptartamento de Matem\'atica, Instituto de Ci\^encias Exatas e Naturais, 78735-910, Rondon\'opolis, MT, Brazil}                           \author{Victor O. Rivelles} \email{rivelles@fma.if.usp.br}
\affiliation{Instituto de F\'\i sica, Universidade de S\~ao Paulo, C.Postal 66318, 05314-970, S\~ao Paulo, SP, Brazil}


\begin{abstract}
We present the transition amplitude for a particle moving in a space with two times and $D$ space dimensions having a $Sp(2,R)$ local symmetry and an $SO(D,2)$ rigid symmetry. It was obtained from the BRST-BFV quantization with a unique gauge choice. We show that by constraining the initial and final points of this amplitude to lie on some hypersurface of the $D+2$ space the resulting amplitude reproduces well known systems in lower dimensions. This work provides an alternative way to derive the effects of 2T physics where all the results come from a single transition amplitude.
 
\end{abstract}


\maketitle



Formulating physics with more than one time leads to unsurmountable problems. However an elegant solution was found for the case of two times (2T) assuming the existence of a $Sp(2,R)$ gauge symmetry, involving a space with two timelike and an arbitrary number of spacelike directions, which reduces the theory effectively to one time after gauge fixing \cite{Bars:1998ph}. The simplest model we can formulate is for a particle moving in $D+2$ dimensions in a space with metric $(-,+,-,+,\dots,+)$ in $D$ spatial and two time dimensions having an $SO(D,2)$ isometry group. It was shown in a series of works that different gauge choices or different choices for the Hamiltonian among the generators of $SO(D,2)$ lead to different systems, like the massless relativistic particle in $D$ spacetime dimension \cite{Bars:1998pc}, a particle in $AdS_{D-n}\times S^n$ \cite{Bars:1998iy}, a massive non-relativistic particle in $D-1$ spatial dimensions \cite{Bars:1998iy}, the harmonic oscillator in $D-2$ spatial dimensions \cite{Bars:1998pc} or  the hydrogen atom in $D-1$ spatial dimensions \cite{Bars:1998pc}. The common characteristic of these systems is that all of them share the original $SO(D,2)$ symmetry usually realized non-linearly in a hidden form. The formulation of 2T physics was also extended to the supersymmetric case \cite{Bars:1998ui}, to curved spaces \cite{Bars:2008sz},  and to field theories \cite{Bars:2000mz} and strings \cite{Bars:1997xb} with the same surprising results that different physical systems emerge out of it \cite{Bars:1998cs}. As stated above the emergence of different systems from the 2T theory is made in different settings. 
Of course it would be desirable to have a formulation where all of its richness could be derived in just one single framework. This is precisely the aim of this paper. The transition amplitude for a particle in 2T physics is found by using the BRST quantization scheme in which only a single gauge choice is done. We will show that amplitudes for different lower dimensional systems are found when we take the initial and final points of the 2T particle in different hypersurfaces of the original $D+2$ space. Our results provide a more physical interpretation of 2T physics. The motion of the 2T particle restricted to different subspaces can be seen as a massless scalar field in Minkowski spacetime in $D$ dimensions or a massive scalar field in $AdS_D$ or any of the systems found in 2T physics. It provides an alternative way to understand 2T theories and strongly corroborates the results found earlier. Our results also have some of the flavor of M-theory in the sense that the restriction to different corners of the $D+2$ space reveals completely different theories in lower dimensions. 

A particle moving in 2T theory has a phase space described by $(X^M, P_M)$, $M=0^\prime,1^\prime,0,1,\dots,D-1$. The gauge symmetries are generated by the three first class constraints $X^2 = X^M X_M, P^2 = P_M P^M$ and $PX = P_M X^M$ whose Poisson bracket algebra generates $Sp(2,R)$ \cite{Bars:1998cs}. Since the BRST-BFV formalism \cite{Fradkin:1975cq} is a powerful method to deal with gauge theories \cite{Rivelles:1995hm} we performed the BRST-BFV quantization  of the 2T particle. The particle is described by its path $X^M(\tau)$ in the 2T space where $\tau$ is its evolution parameter. The action is composed by the usual kinetic term for a particle plus the constraints multiplied by their corresponding Lagrange multipliers $A^a(\tau)$, $a=1,2,3$, which will enforce them. The action is written as 
\begin{equation} \label{1}
	S = \int d\tau \left( P_M \dot{X}^M - A^1 X^2 - A^2 PX - A^3 P^2 \right), 
\end{equation}
where dot means derivative with respect to $\tau$. This action is reparametrization invariant so its canonical Hamiltonian vanishes identically \cite{Bars:1998ph}. According to the BRST-BFV prescription we have to supplement the action (\ref{1}) with ghosts and momenta for the Lagrange multipliers so that the full action is invariant under a nilpotent BRST symmetry. Then a gauge fixing term, which is also BRST invariant, is further added to the action. The gauge choice that was made is  $A^1 = A^2 = 0, \dot{A}^3 = 0$. The evaluation of the path integral representing the transition amplitude for this particle is rather involved and has many subtleties. It will be derived in full details in another publication \cite{new}. Here we will just present and discuss the resulting path integral. The transition amplitude for a particle in $D+2$ dimensions is giving by
\begin{equation} \label{2}
	{\cal A}(X_i, X_f) = \delta(X^2_i) \delta(X^2_f) \int^{+\infty}_{-\infty} d\alpha \,\, \alpha^{D/2-2} \,\, e^{i\alpha X_i X_f},
\end{equation}
where $X_i = X(\tau_i)$ and $X_f = X(\tau_f)$ are the initial and final points of the particle respectively. Notice that it is manifestly invariant under $SO(D,2)$ but not under translations. Also, it does not depend on $\tau_i$ and $\tau_f$  To make sure that we have the right amplitude we can make a few checks. First of all, as just mentioned, it is invariant under the $SO(D,2)$ global symmetry of the classical theory as expected. It also has to obey Schr\H{o}edinger equation $H{\cal A} = -i \partial{\cal A}/\partial\tau$. Since the canonical Hamiltonian vanishes and the amplitude does not depend on $\tau$ this is also verified. Finally it has to satisfy all three constraints \footnote{The equations for a free scalar field in 2T physics \cite{Bars:2006dy} are the same as the ones being discussed here. The context however is different since we are working in a worldline formulation.}. The constraint $X^2=0$ is trivially satisfied for both arguments. The constraint $P^2 {\cal A} = - \partial^2 {\cal A}/\partial X^2 = 0$ requires some manipulations with delta functions but is also satisfied. Finally the last constraint has ordering problems but we find that the amplitude (\ref{2}) satisfies the symmetric ordered constraint $ (X^M \frac{\partial}{\partial X^M} + \frac{\partial}{\partial X^M} X^M) {\cal A} = 0$ \footnote{Notice that the solution to the constraints may not be unique. For instance, multiplying the integrand of (\ref{2}) by a factor of $\epsilon(\alpha)$ gives another solution of the constraint equations which, however, it is not derivable from the BRST-BFV formalism.}. In order to better handle the delta functions we will take the Fourier transform of the amplitude with respect to one of the extra coordinates. Using light cone coordinates $(X^{+^\prime},X^{-^\prime})$ for the sector $(X^{0^\prime},X^{1^\prime})$ we take a Fourier transform with respect to say $X^{+^\prime}$ obtaining
\begin{widetext}
\begin{equation}
	\label{3}
	\tilde{A}(p^{-^\prime}_i, X^{-^\prime}_i, X^\mu_i; p^{-^\prime}_f, X^{-^\prime}_f, X^\mu_f) = \frac{\exp \frac{1}{2} \left( \frac{p^{-^\prime}_i}{X^{-^\prime}_i} (X^\mu_i)^2 + \frac{p^{-^\prime}_f}{X^{-^\prime}_f} (X^\mu_f)^2 \right) }{|X^{-^\prime}_i X^{-^\prime}_f|} \int d\alpha \,\, \alpha^{D/2-2} \exp \left[ \frac{i}{2} \alpha \left( \sqrt{\frac{X^{-^\prime}_i}{X^{-^\prime}_f}} X^\mu_f - \sqrt{\frac{X^{-^\prime}_f}{X^{-^\prime}_i}} X^\mu_i \right)^2 \right],
\end{equation}
\end{widetext}
where $\mu=0,\dots,D-1$ and $p^{-^\prime}$ is the Fourier transformed variable. 

We can now consider what happens when the initial and final points of the particle rest on some hyperplanes of the $D+2$ space. So let us choose $X^{-^\prime} = 1$ and $p^{-^\prime}=0$. We obtain immediately that 
\begin{equation}
	\label{4}
	\tilde{A}(  X^\mu_i; X^\mu_f) = \Delta_F(\Delta X^\mu),
\end{equation}
where $\Delta_F$ is the Feynman propagator for the massless Klein-Gordon field in Minkowski spacetime in $D$ dimensions. So if the initial and final points in the 2T particle amplitude reside in the surface defined by $X^{-^\prime} = 1$ and $p^{-^\prime}=0$ then the 2T particle behaves like the massless relativistic particle in two dimensions less, that is, in $D$ spacetime dimensions. It should be remarked that (\ref{4}) is not the transition amplitude for the relativistic particle computed, for instance, in the BRST formalism. The transition amplitude for the relativistic particle, which satisfies the relativistic particle constraint, is the Schwinger function $\Delta^{(1)}(\Delta X^\mu)$ satisfying the Klein-Gordon equation \cite{Henneaux:1992fc}. As we will see, this is a common feature for relativistic systems.  We get the propagator of some field theory and not the transition amplitude for the associated particle. For the non-relativistic systems we will always get the quantum mechanical transition amplitude. Another remark is that instead of choosing $X^{-^\prime} = 1$ we could have chosen $X^{-^\prime} = \mbox{constant}$. Then the RHS of (\ref{4}) would be multiplied by $1/\mbox{constant}^2$ and the coefficient of the propagator would recall its higher dimensional origin. 

The next corner we are going to look will show us the propagator of a scalar field in $AdS_D$. To this end we will make the change of variables from $(X^0, X^i, X^{D-1}, X^{-^\prime})$ to $(t, x^i, y, R)$, with $i=1,\dots,D-2$, given by
\begin{equation} \label{5}
	X^0 = R t/y, \,\, X^i= R x^i/y, \,\, X^{D-1} = R, \,\, X^{-^\prime} = R/y.
\end{equation}
As we shall see $(t,x^i,y)$ will turn out to be Poincar\'e coordinates for $AdS_D$. With this choice we get  
\begin{equation} \label{6}
	\left( \sqrt{\frac{X^{-^\prime}_i}{X^{-^\prime}_f}} X^\mu_f - \sqrt{\frac{X^{-^\prime}_f}{X^{-^\prime}_i}} X^\mu_i \right)^2 = 2 R_i R_f u,
\end{equation}
where $u = [ (\Delta x^\mu)^2 + y^2_i + y^2_f ]/(2y_iy_f) -1$ is the chordal distance in $AdS_D$. After rescaling the integration variable by $\alpha \rightarrow \alpha/(R_i R_f)$ and choosing the hypersurface $p^{-^\prime} = 0$ and $ y -  R^{D/2} = 0$, we can eliminate $R$ to get for (\ref{3}) (up to numerical factors)
\begin{equation} \label{7}
	\tilde{A}(x^\mu_i, y_i; x^\mu_f, y_f) = \frac{\Gamma(D/2-1)}{(u + i\epsilon)^{D/2-1}}, 
\end{equation}
which is the propagator for a massive scalar field in $AdS_D$ coupled to the background in  a Weyl invariant way \cite{D'Hoker:2002aw}. Its mass is given by $m^2 = - D(D-2)$ and this value is, of course, above the Breitenlohner-Freedman bound. 

We can generalize the former result to a conformally flat spacetime in a slightly different way. Let us choose $p^{-^\prime} = 0$ and $X^{-^\prime} = \exp \sigma(x^\mu)$, and perform the change of variables $ X^\mu = \exp\sigma(x) x^\mu$. After the rescaling of the integration variable by $\alpha \rightarrow \alpha \exp( -\sigma(x_i) -\sigma(x_f) )$ we find that (\ref{3}) reduces to 
\begin{equation} \label{8}
	\tilde{\cal A}(x^\mu_i, x^\mu_f) = \exp\left[{-\frac{D}{2} (\sigma(x_i) + \sigma(x_f))}\right] \Delta_F(\Delta x^\mu),
\end{equation}
where, as before, $\Delta_F$ is the massless propagator of the Klein-Gordon field in flat Minkowski spacetime in $D$ dimensions. Clearly $\exp\sigma(x)$ is a conformal factor. The result (\ref{8}) is proportional to the propagator of the massless Klein-Gordon field conformally coupled to the background \cite{Birrell:1982ix}. There is a factor of $\exp((\sigma(x_i) + \sigma(x_f))$ missing in it so the result (\ref{8}) seems to recall its higher dimensional origin. Alternatively we could have started with $\exp((1-2/D)\sigma(x))$ everywhere instead of $\exp\sigma(x)$ and the right conformal factor would appear in (\ref{8}). Our results for the coupling to $AdS_D$ and to the conformally flat spacetime completely agree with the ones obtained in the context of 2T field theory \cite{Bars:2007kw,Bars:2000mz}. 

Finally we will derive two nonrelativistic systems. The first is the nonrelativistic massive particle in $D-1$ spatial dimensions. To that end we will take a Fourier transform on $X^0_i$ and $X^0_f$ in (\ref{3}), with transformed variables $p^0_i$ and $p^0_f$ respectively. We choose the hypersurface $p^{-^\prime}=0$ and $X^{-^\prime}= 1/\sqrt{m}$ with $m$ constant. After rescaling the integral by $\alpha \rightarrow m\alpha$  and calling $p^0_f = \sqrt{mE}$, with $E$ another constant we get 
\begin{widetext}
\begin{equation} \label{9}
	\tilde{\cal A}(\vec{X}_i, \vec{X}_f, m, E) = m^{d/2}  \int d\alpha \,\, \alpha^{d/2-2} \exp\left[ {\frac{i}{2} ( m\alpha \Delta\vec{X}^2 + \frac{E}{\alpha} )}\right] \delta(p^0_i + \sqrt{mE}) ,
\end{equation}
\end{widetext}
where $d=D-1$. Up to the delta function this is the amplitude for a nonrelativistic particle of mass $m$ and energy $E$ in $d$ spatial dimensions. This result is not so surprising since it could have been obtained directly from the propagator of the massless relativistic particle. The connection between the massless relativistic particle and the massive nonrelativistic particle is well know and it is due to the presence, again, of the $SO(D,2)$ symmetry \cite{Leiva:2003kd}.  

For the last case we will use light-cone coordinates for the $X^\mu$ sector $(X^+, X^-, \vec{X})$. We start by taking the Fourier transform of $X^-_i,X^-_f$ with transformed variables $p^+_i, p^+_f$ respectively. This Fourier transform is used to get two delta functions and one of them can be used to perform the integral on $\alpha$. We now change variables from $(X^+, X^{-^\prime}, p^+, p^{-^\prime})$ to $(t, \tilde{x}^{-^\prime}, \tilde{p}^+, \tilde{p}^{-^\prime})$ according to 
\begin{equation} \label{10}
	X^+ = e^t, \,\, X^{-^\prime} = \tilde{x}^{-^\prime} e^{-t}, \,\, p^+ = \tilde{p}^+ e^t, \,\, p^{-^\prime} = \tilde{p}^{-^\prime} e^{-t},
\end{equation}
and choose the hypersurface defined by $\tilde{x}^{-^\prime} = 1$ and $\tilde{p}^+ + \tilde{p}^{-^\prime} = 0$. The final result is 
\begin{widetext}
\begin{equation} \label{11}
	\tilde{\cal A}(\tilde{p}^+_i, t_i, \vec{X}_i; \tilde{p}^+_f, t_f, \vec{X}_f) = (\tilde{p}^+_f)^{d/2-1} \,\,\, \sinh^{-d/2}\Delta t \,\,\, \exp\left[ \frac{i}{2} \tilde{p}^+_f \frac{\cosh t}{\sinh t} \left( \vec{X}_i^2 + \vec{X}_f^2 - \frac{2 \vec{X}_i \vec{X}_f}{\cosh t} \right) \right] \delta(\tilde{p}^+_i + \tilde{p}^+_f), 
\end{equation}
\end{widetext}
where $d=D-2$. Notice the appearance of hyperbolic functions in (\ref{11}). It is proportional to the transition amplitude for the inverted harmonic oscillator in $d$ spatial dimensions with mass $\tilde{p}^+_f$. The inverted harmonic oscillator has the wrong sign for the potential  but even being unbounded from below it can be quantized in order to study unstable systems \cite{Barton:1984ey}. Notice that to get the amplitude for the ordinary harmonic oscillator we should have to replace $t$ by $it$ in (\ref{10}) but this is not allowed by the change of variables if they are all real. In the usual formulation of 2T physics an ordinary harmonic oscillator is obtained \cite{Bars:1998pc}. Since the parametrization of our hypersurface does not seem to be related to the parametrization used in \cite{Bars:1998pc} it appears that we are looking at a different sector of the 2T theory. 

Finally some comments are in order. If we compare our hypersurfaces with the gauge choices made in the usual formulation of 2T physics \footnote{There are two nice tables in \cite{Bars:2007kw} where all of the cases studied here, except for the harmonic oscillator, are summarized.} we see some similarity. This is to be expected because in the amplitude (\ref{3}) we have explicitly to use the constrains $X^2=0$ which is also done when the gauge is fixed in the usual 2T formulation. So it is natural that some components of $X^M$ and $P^M$ do indeed coincide but the identification is not exact in all cases. It should also be noticed that the inverted harmonic oscillator was the only case with a non vanishing $p^{-^\prime}$ that we were able to analyze. We could not find the hypersurfaces which give rise to other systems like the massive relativistic particle or the hydrogen atom. It seems that the reason is related to the inclusion of the momenta in the definition of the hypersurface. In fact if we make a naive Fourier transformation to get the transition amplitude in phase space we find that it is problematic. Work in this direction is under way. 

It is quite remarkable that a single transition amplitude in $D+2$ dimensions can give rise to several distinct amplitudes in two or three dimensions less. As fully explained in many papers on 2T physics they all share a common $SO(D,2)$ symmetry. This looks like a toy model realization of M-theory. Here we have the emergence of different systems when we fix the initial and final points of the 2T particle to rest on a given subspace of the $D+2$ original space. Even though a 2T formulation of supersymmetry and strings are known we can not leave aside the deep connection of $SO(D,2)$ and the AdS/CFT correspondence. We are planning to extend the formalism presented here to include supersymmetry and strings and try to shed some new light on these matters.

\begin{acknowledgments}
V.O.R. would like to thank Itzhak Bars for helpful discussions and comments. 
The work of J.E.F. was supported in its initial stage by CNPq. 
The work of V.O.R. is supported by CNPq grant No. 304495/2007-7, FAPESP grant No. 2008/05343-5 and PROSUL grant No. 490134/2006-8.
\end{acknowledgments}


\end{document}